# SEARCHES FOR THE ROLE OF SPIN AND POLARIZATION IN GRAVITY: A FIVE-YEAR UPDATE[*]

WEI-TOU NI

*Center for Gravitation and Cosmology, Department of Physics, National Tsing Hua University
Hsinchu, Taiwan 30013, ROC
weitou@gmail.com*

Searches for the role of spin in gravitation dated before the firm establishment of the electron spin in 1925. Since mass and spin, or helicity in the case of zero mass, are the Casimir invariants of the Poincaré group and mass participates in universal gravitation, these searches are natural steps to pursue. In this update, we report on the progress on this topic in the last five years after our last review. We begin with how is Lorentz/Poincaré group in local physics arisen from spacetime structure as seen by photon and matter through experiments/observations. The cosmic verification of the Galileo Equivalence Principle for photons/electromagnetic wave packets (Universality of Propagation in spacetime independent of photon energy and polarization, i.e. nonbirefringence) constrains the spacetime constitutive tensor to high precision to a core metric form with an axion degree and a dilaton degree of freedom. Hughes-Drever-type experiments then constrain this core metric to agree with the matter metric. Thus comes the metric with axion and dilation. In local physics this metric gives the Lorentz/Poincaré covariance. Constraints on axion and dilaton from polarized/unpolarized laboratory/astrophysical/cosmic experiments/observations are presented. In the end, we review the theoretical progress on the issue of gyrogravitational ratio for fundamental particles and the experimental progress on the measurements of possible long range/intermediate range spin-spin, spin-monopole and spin-cosmos interactions

## 1. Introduction

Both electroweak and strong interactions are strongly spin-dependent. In searching for the role of spin in gravitation, we look into the empirical foundations of current theories of gravitation, i.e. general relativity and other relativistic theories of gravity. Relativity sprang out from Maxwell-Lorentz theory of electromagnetism. Therefore we first look into the empirical role of polarization and spin in the gravity-coupling of electromagnetism; we look into photon/electromagnetic wave packet propagation in spacetime. Astrophysical observations and cosmological observations have shown that our early universe evolves according to Hot Big Bang theory. Particle physics experiments have established the Standard Model. ATLAS and CMS experiments in LHC (Large Hadron Collider) have discovered the Higgs particle and pushed the validity of the Standard Model to above the Higgs energy empirically. After the electroweak phase transition around Higgs energy about 100 ps since the Hot Big Bang, the electromagnetism separated out and the Maxwell-Lorentz theory and QED became valid. We look for empirical evidences for gravity-coupling of electromagnetism since then in section 2. In Paper I,[1] we reviewed the experimental and theoretical efforts in searches for the role of spin and polarization in gravity up to the end of 2009. In the present paper, we give a five-year update. In section 3, we review theoretical works on the gyrogravitational effects. In section 4, we review experimental progress on the measurement of long range/intermediate range spin-spin, spin-monopole and spin-cosmos interactions. In section 5, we look into the future.

---

[*]Plenary talk presented in the 21st International Symposium on Spin Physics (Spin2014), 19 October 2014 - 24 October 2014, Peking University, Beijing, China.

## 2. Search for Polarization Effects in the Coupling of Gravity to Electromagnetism

In the genesis of general relativity, there are two important cornerstones: the Einstein Equivalence Principle (EEP) and the metric as the dynamic quantity of gravitation[2] (See also, Ref. [3]). With research activities on cosmology thriving, people have been looking actively for alternative theories of gravity again for more than thirty years. Recent theoretical studies include scalars, pseudoscalars, vectors, metrics, bimetrics, strings, loops, etc. as dynamic quantities of gravity. To find out how metric and other possible fields arises from experiments/observations, we notice that Maxwell-Lorentz electrodynamics can be put into premetric form dependent only on differential structure, not on metric/connection or other geometric structures.[4-9]

### 2.1. *Premetric formulation of electromagnetism*

Maxwell equations for macroscopic/spacetime electrodynamics in terms of independently measurable field strength $F_{kl}$ (*E*, *B*) and excitation (density with weight +1) $H^{ij}$ (*D*, *H*) do not need metric as primitive concept (See, e. g., Hehl and Obukhov [9]):

$$H^{ij}{}_{,j} = -4\pi J^i, \tag{1a}$$
$$e^{ijkl} F_{jk,l} = 0, \tag{1b}$$

with $J^k$ the charge 4-current density and $e^{ijkl}$ the completely anti-symmetric tensor density of weight +1 ($e^{0123} = 1$). We use units with the nominal light velocity parameter $c$ equal to 1. To complete this set of equations, a constitutive relation is needed between the excitation and the field:

$$H^{ij} = \chi^{ijkl} F_{kl}. \tag{2}$$

Both $H^{ij}$ and $F_{kl}$ are antisymmetric, hence $\chi^{ijkl}$ must be antisymmetric in $i$ and $j$, and in $k$ and $l$. Therefore the constitutive tensor density $\chi^{ijkl}$ (with weight +1) has 36 independent components, and can be uniquely decomposed into principal part (P), skewon part (Sk) and axion part (Ax) as given in [9, 10]:

$$\chi^{ijkl} = {}^{(P)}\chi^{ijkl} + {}^{(Sk)}\chi^{ijkl} + {}^{(Ax)}\chi^{ijkl}, \quad (\chi^{ijkl} = -\chi^{jikl} = -\chi^{ijlk}), \tag{3}$$

with

$$^{(P)}\chi^{ijkl} = (1/6)[2(\chi^{ijkl} + \chi^{klij}) - (\chi^{iklj} + \chi^{ljik}) - (\chi^{iljk} + \chi^{jkil})], \tag{4a}$$

$$^{(Ax)}\chi^{ijkl} = \chi^{[ijkl]} = \varphi \, e^{ijkl}, \tag{4b}$$

$$^{(Sk)}\chi^{ijkl} = (1/2) (\chi^{ijkl} - \chi^{klij}). \tag{4c}$$

The principal part has 20 degrees of freedom. The axion part has one degree of freedom. The Hehl-Obukhov-Rubilar skewon part (4c) can be represented as

$$^{(Sk)}\chi^{ijkl} = e^{ijmk} S_m{}^l - e^{ijml} S_m{}^k, \tag{5}$$

with $S_m{}^n$ a traceless tensor of 15 independent degrees of freedom.[9,10] If there is metric, the $S_m{}^n$ can be raised or lowered with this metric; when $S_{mn}$ is symmetric it is called Type I, and when it is antisymmetric it is called Type II.[11] For the skewonless case (i.e., $\chi^{ijkl} = \chi^{klij}$), the Maxwell equations can be derived from the Lagrangian density:

$$L_I = L_I^{(EM)} + L_I^{(EM-P)} + L_I^{(P)} = -(1/(16\pi))\chi^{ijkl} F_{ij} F_{kl} - A_k J^k + L_I^{(P)}, \tag{6}$$

with $L_I^{(EM)} = -(1/(16\pi))\chi^{ijkl} F_{ij} F_{kl}$, $A_k$ the electromagnetic potential guaranteed by (1b), $J^k$ the 4 charge current density and $L_I^{(P)}$ the particle Lagrangian density. The Lagragian density (6) has been used to study the equivalence principles and their empirical foundations in the 1970s and 1980s.[12,13,14] Photon sector of the Standard Model Extension (SME)[15] is contained in the $\chi^{ijkl}$-framework with $L_I^{(EM)}$ of (6).[16] In the Standard Model Supplement (SMS),[17] photon sector is different from but overlaps with the $\chi^{ijkl}$-framework.

In macroscopic medium, the constitutive tensor gives the medium-coupling to electromagnetism; it depends on the (thermodynamic) state of the medium and, in turn, depends on temperature, pressure etc. In gravity, the constitutive tensor (2) gives the gravity-coupling to electromagnetism; it depends on the gravitational field(s) and, in turn, depends on the matter distribution and its state. Now the issue is how to arrive at the metric from the constitutive tensor through experiments/observations. That is, how to build the metric empirically and test the Einstein Equivalence Principle thoroughly. Are there other degrees of freedom to be explored?

Since ordinary energy compared to Planck energy is very small, in this situation we can assume that the gravitational (or spacetime) constitutive relation tensor is a linear and local function of gravitational field(s).

### 2.2. *Wave propagation and dispersion relation*

The sourceless Maxwell equation (1b) is equivalent to the local existence of a 4-potential $A_i$ such that

$$F_{ij} = A_{j,i} - A_{i,j}, \tag{7}$$

with a gauge transformation freedom of adding an arbitrary gradient of a scalar function to $A_i$. The Maxwell equation (1a) in vacuum then becomes

$$(\chi^{ijkl} A_{k,l})_{,j} = 0. \tag{8}$$

Using the derivation rule, we have

$$\chi^{ijkl} A_{k,lj} + \chi^{ijkl}{}_{,j} A_{k,l} = 0. \tag{9}$$

Neglecting $\chi^{ijkl}{}_{,m}$ for slowly varying/nearly homogeneous field/medium, or in the lowest eikonal approximation, (9) becomes

$$\chi^{ijkl} A_{k,lj} = 0. \tag{10}$$

In the weak field or dilute medium, we assume

$$\chi^{ijkl} = \chi^{(0)ijkl} + \chi^{(1)ijkl} + O(2), \tag{11}$$

where O(2) means second order in $\chi^{(1)}$. Since the deviation/violation from the Einstein Equivalence Principle would be small, in the following we assume that

$$\chi^{(0)ijkl} = (1/2) g^{ik} g^{jl} - (1/2) g^{il} g^{kj}, \tag{12}$$

and that $\chi^{(1)ijkl}$ is small compared with $\chi^{(0)ijkl}$. We can then find a locally inertial frame such that $g^{ij}$ becomes the Minkowski metric $\eta^{ij}$ good to the derivative of the metric. To look for wave solutions, we use eikonal approximation and choose $z$-axis in the wave propagation direction so that the solution takes the following form:

$$A = (A_0, A_1, A_2, A_3)\, e^{ikz - i\omega t}. \tag{13}$$

We expand the solution as

$$A_i = [A^{(0)}{}_i + A^{(1)}{}_i + O(2)]\, e^{ikz - i\omega t}. \tag{14}$$

Imposing radiation gauge condition in the zeroth order in the weak field/dilute medium/weak EEP violation approximation, we find the zeroth order solution of (14) and the zeroth order dispersion relation first and then we derive the dispersion relation to first order in Ref. [11]:

$$\omega = k\left[1 + 1/2\,(A_{(1)} + A_{(2)}) \pm 1/2\,((A_{(1)} - A_{(2)})^2 + 4B_{(1)}B_{(2)})^{1/2}\right] + O(2), \quad (15)$$

with

$$A_{(1)} \equiv \chi^{(1)1010} - (\chi^{(1)1013} + \chi^{(1)1310}) + \chi^{(1)1313}, \quad (16a)$$
$$A_{(2)} \equiv \chi^{(1)2020} - (\chi^{(1)2023} + \chi^{(1)2320}) + \chi^{(1)2323}, \quad (16b)$$
$$B_{(1)} \equiv \chi^{(1)1020} - (\chi^{(1)1023} + \chi^{(1)1320}) + \chi^{(1)1323}, \quad (16c)$$
$$B_{(2)} \equiv \chi^{(1)2010} - (\chi^{(1)2013} + \chi^{(1)2310}) + \chi^{(1)2313}. \quad (16d)$$

From (15) the group velocity is

$$v_g = \partial\omega/\partial k = 1 + 1/2\,(A_{(1)} + A_{(2)}) \pm 1/2\,((A_{(1)} - A_{(2)})^2 + 4B_{(1)}B_{(2)})^{1/2} + O(2). \quad (17)$$

We note that $A_{(1)}$ and $A_{(2)}$ *contain only the principal part of* $\chi$; $B_{(1)}$ *and* $B_{(2)}$ *contain only the principal and skewon part of* $\chi$. *The axion part drops out and does not contribute to the dispersion relation in the eikonal approximation*. The principal part $^{(P)}B$ and skewon part $^{(Sk)}B$ of $B_{(1)}$ are as follows:

$$^{(P)}B = (1/2)(B_{(1)} + B_{(2)});\ ^{(Sk)}B = (1/2)(B_{(1)} - B_{(2)}), \quad (18a)$$
$$B_{(1)} = {}^{(P)}B + {}^{(Sk)}B;\ B_{(2)} = {}^{(P)}B - {}^{(Sk)}B. \quad (18b)$$

The quantity under the square root sign is

$$\xi \equiv (A_{(1)} - A_{(2)})^2 + 4B_{(1)}B_{(2)} = (A_{(1)} - A_{(2)})^2 + 4({}^{(P)}B)^2 - 4({}^{(Sk)}B)^2. \quad (19)$$

Depending on the sign or vanishing of $\xi$, we have the following three cases of electromagnetic wave propagation:

(i) $\xi > 0$, $(A_{(1)} - A_{(2)})^2 + 4({}^{(P)}B)^2 > 4({}^{(Sk)}B)^2$: There is birefringence of wave propagation;
(ii) $\xi = 0$, $(A_{(1)} - A_{(2)})^2 + 4({}^{(P)}B)^2 = 4({}^{(Sk)}B)^2$: There are no birefringence and no dissipation/amplification in wave propagation;
(iii) $\xi < 0$, $(A_{(1)} - A_{(2)})^2 + 4({}^{(P)}B)^2 < 4({}^{(Sk)}B)^2$: There is no birefringence, but there are both dissipative and amplifying modes in wave propagation.

In Ref. [11], we have shown that for $B_{(1)} = B_{(2)}$ (i.e., $^{(Sk)}B = 0$), the nonbirefringence condition (Galileo Equivalence Principle for photons/electromagnetic wave packets) for wave propagation in all directions implies the constitutive tensor can be put into the following form:

$$\chi^{ijkl} = {}^{(P)}\chi^{(1)ijkl} + {}^{(Ax)}\chi^{(1)ijkl} + {}^{(SkII)}\chi^{(1)ijkl}$$
$$= \tfrac{1}{2}\,(-h)^{1/2}[h^{ik}h^{jl} - h^{il}h^{kj}]\psi + \varphi e^{ijkl} + \tfrac{1}{2}\,(-\eta)^{1/2}\,(p^{ik}\eta^{jl} - p^{il}\eta^{jk} + \eta^{ik}p^{jl} - \eta^{il}p^{jk}), \quad (20)$$

to first-order in terms of $h^{(1)ij}$, $\psi$, $\varphi$, and $p^{ij}$ with the fields $h^{(1)ij}$, $(\psi - 1)$, $\varphi$, and $p^{ij}$ defined by appropriate expressions of $\chi^{(1)}$s ($h^{ij} \equiv \eta^{ij} + h^{(1)ij}$, $h \equiv \det h_{ij}$). In the skewonless case, the nonbirefringence condition implies that the constitutive tensor is of the form

$$\chi^{ijkl} = {}^{(P)}\chi^{(1)ijkl} + {}^{(A)}\chi^{(1)ijkl} = \tfrac{1}{2}\,(-h)^{1/2}[h^{ik}h^{jl} - h^{il}h^{kj}]\psi + \varphi e^{ijkl}, \tag{21}$$

as reviewed in Paper I.[1]

To derive the influence of the axion field and the dilaton field of the constitutive tensor (21) on the dispersion relation, one needs to keep the second term in equation (9). This has been done for the axion field in references [1, 18-22], and for the joint dilaton field and axion field in Ref. [23]. Near the origin in a local inertial frame, the dispersion relation in dilaton field $\psi$ and axion field $\varphi$ is

$$\omega = k - (i/2)\,\psi^{-1}\,(\psi_{,0} + \psi_{,3}) \pm \psi^{-1}\,(\varphi_{,0} + \varphi_{,3}) + O(2), \tag{22}$$

for plane wave propagating in the *z*-axis direction. The group velocity is $v_g = \partial\omega/\partial k = 1$; there is no birefringence. For plane wave propagating in direction $n^\mu = (n^1, n^2, n^3)$ with $(n^1)^2 + (n^2)^2 + (n^3)^2 = 1$, the solution is

$$\begin{aligned}A(n^\mu) &\equiv (A_0, A_1, A_2, A_3) = (0, \underline{A}_1, \underline{A}_2, \underline{A}_3)\exp(-i\,kn^\mu x_\mu - i\omega t)\\ &= (0, \underline{A}_1, \underline{A}_2, \underline{A}_3)\exp[-ikn^\mu x_\mu - ikt \pm (-i)\psi^{-1}(\varphi_{,0}t - n^\mu\varphi_{,\mu}n_\nu x^\nu) - (1/2)\,\psi^{-1}(\psi_{,0}t + n^\mu\psi_{,\mu}n_\nu x^\nu)],\end{aligned} \tag{23}$$

where $\underline{A}_\mu = \underline{A}^{(0)}{}_\mu + n_\mu n^\nu \underline{A}^{(0)}{}_\nu$ with $\underline{A}^{(0)}{}_1 = \pm i\,\underline{A}^{(0)}{}_2$ and $\underline{A}^{(0)}{}_3 = 0$ $[n_\mu \equiv (-n^1, -n^2, -n^3)]$. There are polarization rotation for linearly polarized light due to axion field gradient, and amplification/attenuation due to dilaton field gradient.

### 2.3. *Empirical constraints on the spacetime constitutive tensor*

Nonbirefringence (no splitting, no retardation) for electromagnetic wave propagation independent of polarization and frequency (energy) in all directions can be formulated as a statement of Galilio Equivalence Principle for photons. However, the complete agreement with EEP for photon sector requires in addition: (i) no polarization rotation; (iii) no amplification/no attenuation in spacetime propagation; (iii) no spectral distortion. With nonbirefringence, any skewonless spacetime constitutive tensor must be of the form (21), hence no spectral distortion. From (23), (ii) and (iii) implies that the dilaton $\psi$ and axion $\varphi$ must be constant, i.e. no varying dilaton field and no varying axion field; the EEP for photon sector is observed; the spacetime constitutive tensor is of metric-induced form. Thus we tie the three observational conditions to EEP and to metric-induced spacetime constitutive tensor in the photon sector. The three observational constraints are reviewed in the following 3 sub-subsections with accuracies summarized in Table I. In section 2.3.4, we discuss the skewonful case.

### 2.3.1. *Birefringence constraint*

Empirically, the nonbirefringence condition is verified by the pulsar signal propagation, the polarization observations on radio galaxies and the gamma ray burst observations.[3,24] The accuracy of verification of the nonbirefringence condition is good up to $10^{-38}$.

### 2.3.2. *Constraints on the cosmic polarization rotation and the cosmic axion field*

From (23), for the right circularly polarized electromagnetic wave, the propagation from a point $P_1$ (4-point) to another point $P_2$ adds a phase of $\alpha = \varphi(P_2) - \varphi(P_1)$ to the wave; for left circularly polarized light, the added phase will be opposite in sign.[18] Linearly polarized electromagnetic wave is a superposition of circularly polarized waves. Its polarization vector will then rotate by an angle $\alpha$. In the global situation, it is the property of (pseudo)scalar field that when we integrate along light (wave) trajectory the total polarization rotation (relative to no $\varphi$-interaction) is again $\alpha = \Delta\varphi = \varphi(P_2) - \varphi(P_1)$ where $\varphi(P_1)$ and $\varphi(P_2)$ are the values of the scalar field at the beginning and end of the wave. The constraints[1,21,25-27] listed on the axion field are from the UV polarization observations of radio galaxies and the CMB polarization observations -- 0.02 for Cosmic Polarization Rotation (CPR) mean value $|<\alpha>|$ and 0.03 for the CPR fluctuations $<(\alpha - <\alpha>)^2>^{1/2}$.

### 2.3.3. *Constraints on the dilaton field and constraints on the unique physical metric*

The amplification/attenuation induced by dilaton is independent of the frequency (energy) and the polarization of electromagnetic waves (photons). From observations, the agreement[28] with and the precise calibration of the cosmic microwave background (CMB) to blackbody radiation constrains the fractional change of dilaton $|\Delta\psi|/\psi$ to less than about $8 \times 10^{-4}$ since the time of the last scattering surface of the CMB.[23] Eötvös-type experiments constrain the fractional variation of dilaton to $\sim 10^{-10}$ $U$ where $U$ is the dimensionless Newtonian potential in the experimental environment.[1] Vessot-Levine redshift experiment and Hughes-Drever-type experiments give further constraints.[1]

Table I. Constraints on the spacetime constitutive tensor $\chi^{ijkl}$ and construction of the spacetime structure (metric + axion field $\varphi$ + dilaton field $\psi$) from experiments/observations in skewonless case (*U*: Newtonian gravitational potential). $g_{ij}$ is the particle metric.

| Experiment | Constraints | Accuracy |
|---|---|---|
| Pulsar Signal Propagation | | $10^{-16}$ |
| Radio Galaxy Observation | $\chi^{ijkl} \to \frac{1}{2}(-h)^{1/2}[h^{ik}h^{jl} - h^{il}h^{kj}]\psi + \varphi e^{ijkl}$ | $10^{-32}$ |
| Gamma Ray Burst (GRB) | | $10^{-38}$ |
| CMB Spectrum Measurement | $\psi \to 1$ | $8 \times 10^{-4}$ |
| Cosmic Polarization Rotation Experiment | $\varphi - \varphi_0 (\equiv \alpha) \to 0$ | $|<\alpha>| < 0.02$, $<(\alpha - <\alpha>)^2>^{1/2} < 0.03$ |
| Eötvös-Dicke-Braginsky Experiments | $\psi \to 1$ <br> $h_{00} \to g_{00}$ | $10^{-10}$ $U$ <br> $10^{-6}$ $U$ |
| Vessot-Levine Redshift Experiment | $h_{00} \to g_{00}$ | $1.4 \times 10^{-4}$ $\Delta U$ |
| Hughes-Drever-type Experiments | $h_{\mu\nu} \to g_{\mu\nu}$ <br> $h_{0\mu} \to g_{0\nu}$ <br> $h_{00} \to g_{00}$ | $10^{-24}$ <br> $10^{-19}$ - $10^{-20}$ <br> $10^{-16}$ |

### 2.3.4. *Constraints on the skewon field and the asymmetric metric*

For metric principal part plus skewon part, we have shown that the Type I skewon part is constrained to < a few × $10^{-35}$ in the weak field/weak EEP violation limit.[11] Type II skewon part is not constrained in the first order.[11] However, in the second order it induces birefringence; the nonbirefringence observations constrain the Type II skewon part to ~ $10^{-19}$.[24] However, an additional nonmetric induced second-order contribution to the principal part constitutive tensor compensates the Type II skewon birefringence and makes it nonbirefringent.[24] This second-order contribution is just the extra piece to the (symmetric) core-metric principal constitutive tensor induced by the antisymmetric part of the asymmetric metric tensor $q^{ij}$ (Table II).[24]

Table II.[24] Various 1st-order and 2nd-order effects in wave propagation on media with the core-metric based constitutive tensors. $^{(P)}\chi^{(c)}$ is the extra contribution due to antisymmetric part of asymmetric metric to the core-metric principal part for canceling the skewon contribution to birefringence/amplification-dissipation.

| Constitutive tensor | Birefringence (in the geometric optics approximation) | Dissipation/ amplification | Spectroscopic distortion | Cosmic polarization rotation |
|---|---|---|---|---|
| Metric: ½ $(-h)^{1/2}[h^{ik} h^{jl} - h^{il} h^{kj}]$ | No | No | No | No |
| Metric + dilaton: ½ $(-h)^{1/2}[h^{ik} h^{jl} - h^{il} h^{kj}]\psi$ | No (to all orders in the field) | Yes (due to dilaton gradient) | No | No |
| Metric + axion: ½ $(-h)^{1/2}[h^{ik} h^{jl} - h^{il} h^{kj}]$ + $\varphi e^{ijkl}$ | No (to all orders in the field) | No | No | Yes (due to axion gradient) |
| Metric + dilaton + axion: ½ $(-h)^{1/2}[h^{ik} h^{jl} - h^{il} h^{kj}]\psi$ + $\varphi e^{ijkl}$ | No (to all orders in the field) | Yes (due to dilaton gradient) | No | Yes (due to axion gradient) |
| Metric + type I skewon | No to first order | Yes | Yes | No |
| Metric + type II skewon | No to first order; yes to 2nd order | No to first order and to 2nd order | No | No |
| Metric + $^{(P)}\chi^{(c)}$+ type II skewon | No to first order; no to 2nd order | No to first order and to 2nd order | No | No |
| Asymmetric metric induced: ½ $(-q)^{1/2}(q^{ik}q^{jl} - q^{il}q^{jk})$ | No (to all orders in the field) | No | No | Yes (due to axion gradient) |

## 3. Gyrogravitational Ratio

Gyrogravitational effect is defined to be the response of an angular momentum in a gravitomagnetic field produced by a gravitating source having a nonzero angular momentum. Ciufolini and E. C. Pavlis[29] have measured and verified this effect with 10-30 % accuracy for the dragging of the orbit plane (orbit angular momentum) of a satellite (LAGEOS) around a rotating planet (earth) predicted for general relativity by Lense and Thirring[30]. Gravity Probe B[31] has measured and verified the dragging of spin angular momentum of a rotating quartz ball predicted by Schiff[32] for general relativity with 19 % accuracy. GP-B experiment has also verified the Second Weak Equivalence Principle (WEP II) for macroscopic rotating bodies to ultra-precision.[33]

Just as in electromagnetism, we can define gyrogravitational factor as the gravitomagnetic moment (response) divided by angular momentum for gravitational interaction. We use macroscopic (spin) angular momentum in GR as standard, its gyrogravitational ratio is 1 by definition. In Ref. [34], we use coordinate transformations among reference frames to study and to understand the Lense-Thirring effect of a Dirac particle. For a Dirac particle, the wave-function transformation operator from an inertial frame to a moving accelerated frame is obtained. According to equivalence principle, this gives the gravitational coupling to a Dirac particle. From this, the Dirac wave function is solved and its change of polarization gives the gyrogravitational ratio 1 from the first-order gravitational effects. In Teryaev's talk on Spin-gravity Interactions and Equivalence Principle, he has reported his work with Obukhov and Silenko[35] on the direct calculation of the response of the spin of a Dirac particle in gravitomagnetic field and showed that it is the same as the response of a macroscopic spin angular momentum in general relativity (See, also, Tseng [36]). Randono has showed that the active frame-dragging of a polarized Dirac particle is the same as that of a macroscopic body with equal angular momentum.[37] All these results are consistent with EEP and the principle of action-equal-to-reaction. However, these findings do not preclude that the gyrogravitational ratio to be different from 1 in various different theories of gravity, notably torsion theories and Poincaré gauge theories.

What would be the gyrogravitational ratios of actual elementary particles? If they differ from one, they will definitely reveal some inner gravitational structures of elementary particles, just as different gyromagnetic ratios reveal inner electromagnetic structures of elementary particles. These findings would then give clues to the microscopic origin of gravity.

Promising methods to measure particle gyrogravitational ratio include:[1] (i) using spin-polarized bodies (e.g. polarized solid He$^3$, Dy-Fe, Ho-Fe, or other compounds) instead of rotating gyros in a GP-B type experiment to measure the gyrogravitational ratio of various substances; (ii) atom interferometry; (iii) nuclear spin gyroscopy; (iv) superfluid He$^3$ gyrometry. Notably, there have been great developments in atom interferometry[38] and nuclear gyroscopy.[39] However, to measure particle gyrogravitational ratios the precision is still short by several orders and more developments are required.

## 4. Search for Long Range /Intermediate Range Spin-Spin, Spin-Monopole and Spin-Cosmos Interactions

### 4.1. *Spin-spin experiments*

Geomagnetic field induces electron polarization within the Earth. Hunter et al.[40] estimated that there are on the order of $10^{42}$ polarized electrons in the Earth compared to $\sim 10^{25}$ polarized electrons in a typical laboratory. For spin-spin interaction, there is an improvement in constraining the coupling strength of the intermediate vector boson in the range greater than about 1 km.[40]

### 4.2. *Spin-monopole Experiments*

In Paper I, we have used axion-like interaction Hamiltonian

$$H_{int} = [\hbar(g_s g_p)/8\pi mc] (1/\lambda r + 1/r^2) exp(-r/\lambda) \, \boldsymbol{\sigma} \cdot \hat{\mathbf{r}}, \tag{24}$$

to discuss the experimental constraints on the dimensionless coupling $g_s g_p/\hbar c$ between polarized (electron) and unpolarized (nucleon) particles. In (24), $\lambda$ is the range of the interaction, $g_s$ and $g_p$ are the coupling constants of vertices at the polarized and unpolarized particles, $m$ is the mass of the polarized particle and $\boldsymbol{\sigma}$ is Pauli matrix 3-vector. Hoedll et al.[41] have pushed the constraint to shorter range by about one order of magnitude since our last review. In this update, we see also good progress in the measurement of spin-monopole coupling between polarized neutrons and unpolarized nucleons.[42-44] Tullney et al. obtained the best limit on this coupling for force ranges between $3 \times 10^{-4}$ m and 0.1 m

### 4.3. *Spin-cosmos experiments*

For the analysis of spin-cosmos experiments for elementary particles, one usually uses the following Hamiltonian:

$$H_{cosmic} = C_1 \sigma_1 + C_2 \sigma_2 + C_3 \sigma_3, \tag{25}$$

in the cosmic frame of reference for spin half particle with $C$'s constants and $\sigma$'s the Pauli spin matrices (see, e.g. [45] or Paper I). The best constraint now is on bound neutron from a free-spin-precession $^3$He-$^{129}$Xe comagnetometer experiment performed by Allmendinger et al.[39] The experiment measured the free precession of nuclear spin polarized $^3$He and $^{129}$Xe atoms in a homogeneous magnetic guiding field of about 400 nT. As the laboratory rotates with respect to distant stars, Allmendinger et al. looked for a sidereal modulation of the Larmor frequencies of the collocated spin samples due to (25) and obtained an upper limit of $8.4 \times 10^{-34}$ GeV (68% C.L.) on the equatorial component $C^n_\perp$ for neutron. This constraint is more stringent by $3.7 \times 10^4$ fold than the limit on that for electron.[46] Using a $^3$He-K co-magnetometer, Brown et al.[47] constrained $C^p_\perp$ for the proton to be less than $6 \times 10^{-32}$ GeV.

### 5. Outlook

Polarization and spin are important in verifying Galileo Equivalence Principle and Einstein Equivalence Principle which are important cornerstones of general relativity and metric theories of gravity. General relativity and relativistic theories of gravity are bases for modern cosmology. It is not surprising that cosmological observations on polarization phenomena become the ultimate test ground of the equivalence principles, especially for the photon sector. Some of the dispersion relation tests are reaching second order in the ratio of Higgs boson mass and Planck mass. Ultra-precise laboratory experiments are reaching ground in advancing constraints on various (semi-)long-range spin interactions. Sooner or later, experimental efforts will reach the precision of measuring the gyrogravitational ratios of elementary particles. All these developments may facilitate ways to explore the origins of gravity.

**Acknowledgments**

We would like to thank F. Allmendinger, C. Fu, H. Gao, B.-Q. Ma, Yu. N. Obukhov, O. Teryaev, K. Tullney for helpful discussions. We would also like to thank the National Science Council (Grant No. NSC102-2112-M-007-019) and the National Center for Theoretical Sciences (NCTS) for supporting this work in part.

piece dropped, i.e. $(k_F)^{\kappa\lambda\mu\nu} = \chi^{\kappa\lambda\mu\nu} - (1/2)(\eta^{\kappa\mu}\eta^{\lambda\nu} - \eta^{\kappa\nu}\eta^{\lambda\mu})$. The CPT-odd part $(k_{AF})^{\kappa}$ also has constant components which correspond to the derivatives of axion $\varphi,^{\kappa}$ when specialized to constant values.